\documentstyle{article}
\begin{document}

{\bf {
\centerline{ THE QUANTIZATION OF GRAVITY.}
\centerline{             DYNAMIC APPROACH.}}}

\centerline{ }

\centerline{ \it S.N.Vergeles. \footnote{e-mail:vergeles@itp.ac.ru}}
\vspace{3mm}
\centerline { {\it Russian Academy of Sciences. } }
\centerline { {\it The Landau Institute for Theoretical Physics, Moscow,
 Russia.}}
\centerline { }
\centerline { }

\centerline{\bf {Abstract.}}
\centerline{ }

  \parbox[b]{135mm}{
\baselineskip=14pt
 { \hspace{5mm}
 On the basis of dynamic quantization method we build in this paper
 a new mathematically correct quantization scheme of gravity.
 In the frame of this scheme we develop a canonical formalism in
 tetrad-connection variables in 4-D theory of pure gravity.
 In this formalism the regularized quantized fields corresponding
 to the classical tetrad and connection fields are constructed.  It is
 shown, that the regularized fields satisfy to general covariant equations
 of motion, which have the classical form. In order to solve these equations
 the iterative procedure is offered.}}
\vspace {2mm}

\large {
\baselineskip=18pt

\centerline{\bf {1. Introduction.}}
\centerline{ }

  In this paper we state the mathematically correct quantization scheme
of gravity in 4-dimensional space-time. The basis of the scheme is the
dynamic quantization method. The dynamic quantization method was already
successfully applied to gravity interacting with Dirac field in
2+1-dimensional space-time.  The regularization conserving general
 covariance of the theory was carried out and perturbation theory ( PT )
 was constructed [1,2].  The idea of the  dynamic quantization method is
 based on the Dirac theory of quantization of systems with constraints and
in particular of general covariant systems.

  Describe briefly the dynamic quantization method. As is known [3], the
hamiltonian in general covariant theories is arbitrary linear
combination of the first class constraints $\chi_ {\alpha} $. If $ | \,
{\cal M} \,\rangle$ is any physical state, then $\chi_ {\alpha} | \, {\cal
M} \,\rangle=0 $. Let \ $ | \, {\cal N} \,\rangle=a^ + | \, {\cal M}
\,\rangle$ be other physical state and \ $a^ + $ \ be some
operator of creation type. As the hamiltonian annuls all physical states,
then can assume that \ $ [\chi_ {\alpha},\, a^ +] =0$ \ .  There is an
infinite number of creation and annihilation type
 operators \ $a^ + _N$,   \ $a_N$ \ which transfer
  one physical states in others and exhausting
all local physical degrees of freedom of the system. The all
 operators \ $\{a_N,\, a^ + _N\} $ \ are conserved, since
  they commute with the
hamiltonian. This imply that any set of pairs of the operators \ $\{a_N,\,
a^ + _N\} ^ {\prime\prime} $ \ can be considered as a set of the second
class constraints in the Dirac sense [3]. This fact gives the
following possibility for regularization in considered theory:
the regularization of the system is made by imposing
of an infinite set of second class constraints \\
$$
a_N=0, \qquad \ a^+_N=0 \,, \qquad \ \ | N | > N_0 \eqno (1.1)
$$
Thus, in the theory remains
only final number of degrees of freedom, corresponding to
 the operators \ $a^ + _N$ \ and \ $a_N$ \ with \ $ | N | < N_0$ \ .
 The final set of the
remained operators \ $\{a_N,\, a^ + _N\}^{\prime} $ \ corresponds to the
set of physical states which describe enough completely the investigated
system. As a result a Poisson bracket is replaced by the corresponding
Dirac bracket.  It is critically important that under such regularization
the equations of motion do not change its classical form (see section 4).
 This fact imply that the general covariance in regularized
theory is conserved. Moreover, retaining a "small" number of physical
degrees of freedom and states in the theory, we obtain a new possibility of
development of PT in the number of the remained physical degrees of freedom.

Not concerning the complete review of other directions of canonical
quantization of gravity, we pay attention to
results of one of roughly developed schools, which can be presented by works
[4] (see the references there).  Within the framework of developed by the
authors of [4] technic the principal possibility of canonical
 nonperturbative quantization of gravity is elaborated: the physical states
of theory are described, they form the normalized space;  in this space the
construction of the linear operators, to the number of which the first
  class constraints or hamiltonian belong, is carried out;  the problem of
construction of physical states annuling the hamiltonian is solved; the
commutational relations between the first class constraints do not contain
undesirable Schwinger terms.  In our opinion these results by an indirect
way corroborate our method, causing the same general results.

The method of dynamic quantization is constructed on principles
of canonical quantum theory. Therefore for its application it is
necessary previously to develope the apparatus of classical hamiltonian
mechanics. This problem is solved in Section 2. In Section 3 the formal
construction of a quantum theory is performed. Since in theory of gravity
the formal quantization is mathematiucally not correct, the account in this
Section has heuristic character. In Section 4 the successive logically and
mathematiucally correct quantum theory of gravity is built. In Section 5
the perturbation theory is developed.

\centerline{ }
\centerline{\bf {2. Canonical formalism.}}
\centerline{ }

  We shall consider the first order vierbein action of pure gravity in
4-D space-time:
$$
A=-\frac {1} {8\kappa^2} \int\,
d^4x\,\varepsilon^ {\mu\nu\lambda\rho} \varepsilon_ {abcd} e^c_ {\lambda}
e^d_{\rho} \, R^ {ab} _{\mu\nu} = \int\, dx^0\, {\cal L},
$$
$$
R^ {ab} _
{\mu\nu} =\partial_ {\mu} \omega^ {ab} _ {\nu} - \partial_ {\nu} \omega^
{ab} _ {\mu} + \omega^a_ {\mu g} \omega^ {gb} _ {\nu} - \omega^a_ {\nu g}
\omega^ {gb} _ {\mu} \eqno (2.1)
$$
Here \ $e^a_ {\mu} $ \ are tetrads, so
that $g_ {\mu\nu} =\eta_ {ab} \, e^a_ {\mu} e^b_ {\nu} $ is a metric tensor
in local coordinates $x^ {\mu} = (x^0,\, x^i), \ \mu,\nu,\ldots=0,1,2,3$
are coordinate indexes. The \ $a, b, c,\ldots=0,1,2,3$ are local Lorentz
indexes, \ $\eta_ {ab} =diag (1,-1,-1,-1) $ \ is the Lorentz
 metrics and \ $\omega^ {ab} _ {\mu} =-\omega^ {ba}_{\mu} $ \ is the
connection in orthonormal basis \ $e^a_{\mu} $ \ and the covariant
derivative of the vector \ $\xi^a=e^a_ {\mu} \xi^ {\mu} $ \ is of the form
$$
\nabla_ {\mu}\xi^a= \partial_ {\mu} \xi^a + \omega^ {ab} _ {\mu} \xi_b
$$
Further, let \ $\varepsilon^ {\mu\nu\lambda\rho} \ (\,\varepsilon^ {0123}
=1) $ \ and \ $\varepsilon_ {abcd} \, (\,\varepsilon_ {0123} =1) $ \ be
completely antisymmetric pseudotensors, referred to coordinate and Lorentz
basises, accordingly.

 Now begin the construction of canonical formalism for action (2.1) in the
form convenient for us. This problem was repeatedly considered in
canonically-conjugated variables \ $\omega^ {ab} _i$ \ and
$$
{\cal P} ^i_
{ab} \equiv \frac {\partial {\cal L}} {\partial {\dot { \omega}} ^ {ab} _i}
= - (2\kappa^2) ^ {-1} \,\varepsilon_ {ijk} \,\varepsilon_ {abcd} \,
e^c_je^d_k \eqno (2.2)
$$
( See, for example, [5]). The point on top
means derivative in time \ $t=x^0$ \ . However, since there are
the second class constraints in the system (2.1) the Poisson brackets
in variables \ $\{\omega^ {ab} _i,\, {\cal P} ^i_ {ab} \} $ \ appear rather
complicated. Moreover, the equations of motion in these variables are
enough bulky. On the other hand, in variables \ $\{\omega^ {ab} _i,\,
e^c_j\,\} $ \ as equations of motion, so the constraints look as much as
possible simply. This circumstance is extremely important for us, since in
the dynamic quantization methodof the equations of motion play a main role
in principal calculations.  The
 tetrad-connection variables are prefereble
also because just in these variables the property of supersymmetry in
supergravity is formulated.  Thus the development of hamiltonian formalism
in variables \ $\{\omega^ {ab} _i,\, e^c_j\,\} $ \, having direct  physical
sense seems for us more rational.

By definition
$$
{ \cal L} =
\frac {\partial {\cal L}} {\partial {\dot {q}}} \, {\dot {q}} -H \,
$$
where \ $q$ \ are some generalized coordinates. In our case the hamiltonian
is of the form
$$
H=\int\, d^3x\,\{-\frac {1} {2} \omega^ {ab} _0\chi_
{ab} + \frac {1} {2\kappa^2} e^c_0\phi_c\,\} \ ,
$$
$$
\chi_ {ab} =\frac {1}
{\kappa^2} \varepsilon_ {abcd} \varepsilon_ {ijk} e^c_i\nabla_je^d_k \ ,
\qquad \phi_c=\frac {1} {2} \varepsilon_ {abcd} \varepsilon_ {ijk} e^d_kR^
{ab} _ {ij} \eqno (2.3)
$$
Here \ $\nabla_ {\mu} e^c_ {\nu} =\partial_ {\mu} e^c_ {\nu} + \omega^ {cb} _
  {\mu} e_ {b\nu}, \ \omega^ {ab} _0$ \ and \ $e^c_0$ \ are arbitrary
functions, playing a role of Lagrange multipliers. Let us expresse
the action (2.1) in the form
$$
A=-\int\, d^4x\,\frac {1} {4\kappa^2} \,
\varepsilon_ {abcd} \varepsilon_ {ijk} e^c_je^d_k {\dot {\omega}} ^ {ab}
_j- \int\, dt\, H \eqno (2.4)
$$
From the condition \ $\delta A=0$ \ we find
two equations relatively to \ $ {\dot {\omega}} ^ {ab} _i$ \ and \ $ {\dot
e} ^c_j$ \ :
$$
\frac {1} {2\kappa^2} \,\varepsilon_ {abcd} \varepsilon_
{ijk} e^d_k {\dot {\omega}} ^ {ab} _i + \frac {\partial H} {\partial e^c_j}
=0 \  , \eqno (2.5)
$$
$$
\frac {1} {2\kappa^2} \,\varepsilon_ {abcd}
\varepsilon_ {ijk} {\dot {e}} ^c_je^d_k- \frac {\partial H} {\partial
\omega^ {ab} _i} =0 \eqno (2.6)
$$
Eq.(2.6) has the solution only under additional conditions
$$
\lambda^ {ij} = (g^ {ik} \varepsilon_ {jlm} + g^
{jk} \varepsilon_ {ilm} \,) \, e_{ak} \nabla_le^a_m=0 \ , \eqno (2.7)
$$
where \ $g^ {ik} g_ {kj} =\delta^i_j$ \ . It is necessary to
consider Eq. (2.7) as the second class constraints. It is seen from the
fact that the six equations (2.7) in each point \ $x$ \ reduce the number
 of independent variables \ $\omega^ {ab} _i (x) $ \ from eighteen up to
twelve. The number of independent variables \ $e^a_i (x) $ \ is also equal
to twelve. From Eq.(2.6) under conditions (2.7) we find:
$$
\nabla_0e^a_i=\nabla_ie^a_0 \eqno (2.8)
$$
Besides the constraints \ $\chi_ {ab} \approx 0$ \ and Eq.(2.7) give
$$
 \nabla_ie^a_j-\nabla_je^a_i=0 \ (mod \ \chi_ {ab}) \eqno (2.9)
$$
As is known, the connection is expressed unambiguously from Eq. (2.8)
and (2.9) through tetrads and its
derivatives. Eq.(2.5) determines the quantity \ $ {\dot {\omega}} ^ {ab}
_i$ \ up to the term \ $\varepsilon_ {jkl} e^a_ke^b_ls_ {ij}, \ s_
{ij} =s_ {ji} $ \ . It is possible to find equations of motion with
requirement of conservation for constraints (2.7). Write out
the equations of motion for the connection:
$$
R^{ab}_{0i}
=\frac {1} {2} \varepsilon_c {} ^ {dab} \varepsilon_ {jkl} {\tilde {e}} _
{0d} e_ {if} e_ {0g} e^ {(f} _lR^ {c) g} _ {jk} -
$$
$$
\frac {1} {2} \,
({\tilde {e}} ^ {[a} _0e^ {b]} _le^c_i + \frac {1} {2} \, e^ {[a} _le^ {b]}
 _i {\tilde {e}} ^c_0\,) \cdot\varepsilon_ {cdfg} \varepsilon_ {jkl}
 e^d_0R^ {fg} _ {jk} \ Mod (\chi_ {ab},\,\phi_c\,) \eqno (2.10)
$$
 Here and further \ $ {\tilde{e}}_{0a} =-g^ {-1} \varepsilon_ {abcd}
e^b_1e^c_2e^d_3, \ g=det\, g_{ij}$ \ and \ $ (a\, b) $ \
 (or \ $ [a\, b] $ \ ) means symmetrization (antisymmetrization)
relative to the pair of indexes in the brackets.

Note that Eq. (2.10) and the constraints \ $\phi_c=0$ \ are contained in
equations
$$
\varepsilon^ {\mu\nu\lambda\rho}
\varepsilon_ {abcd} e^c_ {\lambda} R^ {ab} _ {\mu\nu} =0 \eqno (2.11)
$$

 According
to definition of Poisson bracket (PB) the equations of motion
(2.8), (2.10) can be written in the form
$$
{ \dot {A}} = [A,\, H\,]
$$
Here \ $ [\ldots,\ldots] $ \ designates Poisson bracket of the
 quantities \ $A$ \ and \ $H$ \ and \ $A$ \ is any function of dynamic
variables.

 Eqs (2.5) and (2.6) and also the conditions
 \ $ [e^a_i (x),\,\lambda^ {jk} (y) \,] =0$ \ determine unambiguously the
following:
$$
[ e^a_i (x),\, e^b_j (y)] =0 \  , \eqno (2.12)
$$
$$
[ \omega^ {bc} _j (x),\, e^a_i (y)] =\delta (x-y) \,\kappa^2\,
\{\,2\, {\tilde e} ^ {[b} _0\, e^ {c]} _i\, e^a_j +
$$
$$
+ e^{[b} _ie^ {c]} _j\, {\tilde e} ^a_0 +
e^ {[b} _j\, {\tilde e} ^ {c]} _0\, e^a_i\} (x), \eqno (2.13)
$$

   To find  connection-connection Poisson brackets more
long calculations are required. Since further the  connection-connection
Poisson brackets do not used obviously, they are not here written out.

For completion of the description of hamiltonian formalism we shall show,
that under conditions (2.7) the
 quantities \ $\chi_ {ab} $ \ and \ $\phi_c$ \ are the first
  class constraints.

From (2.3), (2.8) and
(2.10) follows, that quantities \ $\chi_{ab}(x) $ \ generate the local
Lorentz transformations in point \ $x$ \ . In
particular
$$
[ \chi_ {ab} (x),\, e^c_i (y) \,] =
-\delta (x-y) (\,\delta^c_ae_ {bi} -\delta^c_be_ {ai} \,) (x),
$$
$$
[ \chi_ {ab} (x),\,\omega^ {cd} _i (y) \,] =
-2\,\partial_i\delta (x-y) \,\delta^ {[c} _a\delta^ {d]} _b-
$$
$$
2\,\delta (x-y) \,\{\delta^ {[c} _a\omega_b {} ^ {d]} _i +
\delta^ {[d} _a\omega^ {c]} _i {} ^b\,\} \, (x)
$$
Any Lorentz-tensor quantity has the Poisson bracket with the quantity
\ $\chi_ {ab} $ \ similar to the written out. Therefore the equation of
motion for \ $\chi_ {ab} $ \ is of the form
$$
{\dot {\chi}} ^ {ab}
=2\,\omega^ {[a} _0 {} _c\chi^ {b] c} - \frac {1} {\kappa^2} \, e^ {[a}
_0\phi^ {b]}
$$
Thus
$$ \partial_ {\mu} \chi^ {ab} =0 \ (mod \ \chi_
{ab},\phi_c\,) \eqno (2.14)
$$

Differentiating in space-time coordinates the curvature tensor (2.1), we
find
$$
\nabla_ {\lambda} R^ {ab} _ {\mu\nu} + \nabla_ {\mu} R^ {ab} _
{\nu\lambda} + \nabla_ {\nu} R^ {ab} _ {\lambda\mu} =0 \ , \eqno (2.15a) $$
 where
 $$
\nabla_ {\lambda} R^ {ab} _ {\mu\nu} =\partial_ {\lambda} R^ {ab} _
{\mu\nu} + \omega^a_ {\lambda c} R^ {cb} _ {\mu\nu} + \omega^b_ {\lambda c}
R^ {ac} _ {\mu\nu}
$$

 Further, differentiating Eqs (2.8) and (2.9) and using (2.14) we
obtain
$$
R^ {ab} _ {\mu\nu} e_ {b\lambda} + R^ {ab} _ {\nu\lambda} e_
{b\mu} + R^ {ab} _ {\lambda\mu} e_ {b\nu} =0
\ (mod \ \chi_ {ab},\,\phi_c\,) \ \eqno (2.15b)
$$

 The relations (2.15) mean that Bianchi identities in
 canonical formalism are valid.

  We state now, that
$$
{ \dot {\phi}} _c=0 \ ( Mod \ \chi_ {ab},\,\phi_c\,) \ \eqno (2.16)
$$
  The equations (2.14) and (2.16) mean that the constraints \ $\chi_
  {ab},\,\phi_c$ \ are the first class constraints.

  Let us show the correctness of Eq. (2.16). From definition of quantities
  \ $\phi_a$ \ we have
  $$
   \nabla_0\phi_a\sim \varepsilon_ {abcd}
  \varepsilon_ {ijk} \, (\,\nabla_0e^b_i\cdot R^ {cd} _ {jk} +
e^b_i\nabla_0R^ {cd} _ {jk} \,) =
$$
$$
 \varepsilon_ {abcd} \varepsilon_
 {ijk} \, [\nabla_ie^b_0\cdot R^ {cd} _ {jk} - e^b_i\, (\,\nabla_jR^ {cd} _
 {k0} + \nabla_kR^ {cd} _ {0j} \,) \,] \eqno (2.17)
  $$
 The last equality is
based on Eq.(2.8) and identity (2.15a). Using (2.9) and (2.15), we
represent the right hand side of (2.17) up to terms proportional to the
constraints \ $\chi_ {ab} $ \ or \ $\phi_c$ \, as
follows:
$$
\nabla_i\, [\, \varepsilon_ {abcd} \varepsilon_ {ijk} \, (\,
e^b_0R^ {cd} _ {jk} - 2\, e^b_jR^ {cd} _ {0k} \,) \,] \ (mod \ \chi_
{ab},\,\phi_c\,) \ \eqno (2.18)
$$
  The quantity in square brackets in (2.18) can be written in the form
   \ $\frac {\delta {\cal L}} {\delta e^a_i} $ \ which is equal to zero
   due to equations of motion. Thus
   equality (2.16) is proven.

\centerline{ }
\centerline{\bf {3. Formal quantization.}}
\centerline{ }

 Passing from the classical mechanics to quantum one we must
replace classical Poisson brackets by quantum coommutation relations.
It is usually assumed that quantum Poisson bracket for fundamental
variables differ from classical one only by a multiplier \ $i (\hbar=1)$ \,
which in
our case stands in right hand side of Eqs (2.12) and (2.13).
 The Heisenberg equations \ $i\, {\dot A} = [\, A,\, H\,] $ \ for
 variables \ $e^a_i$ \ and \ $\omega^ {ab} _i$ \ save its classical form
up to operators ordering.

From commutators (2.12), (2.13) it follows,
that the set of variables \ $\{e^a_i (x) \} $ \ is a complete set of
mutually commuting variables. The possible values of these
variables satisfy to the conditions
$$
 -\infty < e^a_i (x) < + \infty
$$

  We shall write out the formula for operator of connection.
As in this Section a formal quantum theory is considered,
the problem of correct ordering of operators is neglected here.
Using classical PB (2.12) and Eq.
(2.7) we find:
$$
\omega^ {bc} _j (x) =
\kappa^2\,
\{\,2 {\tilde e} ^ {[b} _0\, e^ {c]} _i\, e^a_j +
e^ {[b} _ie^ {c]} _j\, {\tilde e} ^a_0 +
e^ {[b} _j\, {\tilde e} ^ {c]} _0\, e^a_i\} \,\pi^i_a (x) +
$$
$$
+ (2g) ^ {-1} \varepsilon_ {klm} \varepsilon_ {inp} e^b_ne^c_p\,
\partial_le^d_m\cdot
\{\, g_ {ij} e_ {dk} -g_ {jk} e_ {di} -g_ {ik} e_ {dj} \,\}     \eqno (3.1)
$$

Here the field \ $\pi^i_a (x) $ \ is defined by the formulae
$$
[\,\pi^i_a (x),\, \pi^j_b(y) \,]=0 , \qquad
[ \,\pi^i_a (x),\, e^b_j (y) \,] =i\,\delta^i_j\delta^b_a\delta\, (x-y)
 \eqno (3.2)
 $$

 Pass now to the problem of finding of the conserved
 operators, which exshaust the all local physical degrees of freedom.
 The conserved operators in general covariant theories can be
 named also as gauge-invariant operators. The problem of construction
 of gauge-invariant operators
 is resolved much more convenient by an axiomatic approach similar
 to the case of 3-space-time theory of gravity [2].

 Let us introduce the following natural assumptions or axioms
  about the structure of the physical space of states \ $F$\ .
\centerline{ }

{ \underbar { Axiom 1. } { \it { All states of the theory, having the
physical sense, are obtained from the ground state }} \ $\vert
\,0\,\rangle$ \ { \it { with the help of the creation operators }}
 \ $a^ + _N$ \ :
 $$
 \vert \, n_1,\,
 N_1;\ldots;\, n_s,\, N_s\,\rangle= ( n_1!\cdot\ldots\cdot n_s!\, ) ^ {
 -\frac { 1 } { 2 }} \cdot ( a^ + _ { N_1 } ) ^ { n_1 } \cdot\ldots\cdot (
 a^ + _ { N_s } ) ^ { n_s } \,\vert \,0\,\rangle \ ,
 $$
 $$
 a_N\,\vert\,0\,\rangle=0                              \eqno (3.3)
 $$
 { \it
 { The states (3.3) form orthonormal basis of the physical states space
  \ $F$ \ of the theory. }} \
\centerline{ }

The numbers \ $n_1,\ldots, n_s$ \ take the natural valuess and
are called by occupation numbers. \\
\centerline { }

{ \underbar {Axiom 2.}}
{ \it {The states (3.3) satisfy to the conditions:}}
$$
\chi_ {ab} (x) \,\vert \, n_1,\, N_1;\ldots;\, n_s,\, N_s\,\rangle=0,
$$
$$
\phi_a (x) \,\vert \, n_1,\, N_1;\ldots;\, n_s,\, N_s\,\rangle=0
 \eqno (3.4)
$$
\centerline {}

{ \underbar {Axiom 3.}}
{ \it {The state}}
 \ $e^a_i (x) \,\vert\, n_1, N_1;\ldots; n_s, N_s\,\rangle$ \ {\it {contains
 a superposition of {\bf all} states for which one of occupation number
  differs per unit from occupation number of the state (3.3) and
  the rest coincide with corresponding occupation numbers of the state
  (3.3). }} \\
\centerline { }

 The operators \ $a^ + _N$ \ and their Hermithian conjugated \ $a_N$
  \ have usual commutational properties:
$$
[ \, a_N,\, a_M\,] =0, \qquad [a_N,\, a^ + _M\,] =\delta_ {NM} \eqno (3.5)
$$
For the case of the compact spaces one can consider, that index \ $N$ \,
numbering creation and annihilation  operators  belongs to the discret
finite dimensional lattice. In the space of indexes \ $N$ \ the norm can be
easily introduced.

From Eqs. (3.3) and (3.4) it follows, that
\ $ [H,\, a^ + _N\,] \approx 0, \qquad [H,\, a_N\,] \approx 0 \, $ \ .
We shall accept more strong conditions:
$$ [H,\, a^ + _N\,] =0, \qquad [H,\, a_N\,] =0 \ ,
\eqno (3.6)
$$
where the operator \ $H$ \ is given by (2.3).

Thus interesting for us gauge-invariant operators
are formally indicated. This is the set of the
annihilation and creation operators
\ $\{a_N,\, a^ + _N\} $ \,
exshausting the all local physical degrees of freedom of the system.
We pay attention
on the fact that commutational relations (CR) (3.6) are in
agreement with general covariance of the theory. In
 other theories there is no the set of the operators with properties
 (3.6) exshausting  physical degrees of freedom of the system.

Give some consequences from Axioms 1-3.

Let \ $\vert \, N\,\rangle=a^ + _N\,\vert\,0\,\rangle$ \ .
  From Axiom 3 follows, that
$$
e^a_i (x) \,\vert\, N\,\rangle=\frac {1} {\sqrt 2}
e^a_ {Ni} (x) \,\vert\,0\,\rangle + \vert\, N;\, e^a_i (x) \,\rangle,
$$
$$ \langle\,0\,\vert\, N;\, e^a_i (x) \,\rangle=0, \eqno (3.7)
$$
  The fields \ $e^a_ {Ni} (x) $ \ are linearly independent and by definition
  $$
  [\, e^a_ {Ni} (x),\, a_M\,] =0, \qquad
 [\, e^a_ {Ni} (x),\, a^ + _M\,] =0   \eqno (3.8)
 $$
 Since the field \ $e^a_i (x) $ \  is Hermithian there is the following
 expansion:
 $$
 e^a_i (x) =\frac {1} {\sqrt 2} \,\sum_N
  (a_N\, e^a_ {Ni} (x) + a^ + _N\, {\bar e} ^a_ {Ni} (x) \,) +
  {\tilde e} ^a_i (x),
	\eqno (3.9)
$$
The field \ $ {\tilde e} ^a_i (x) $ \ does not contain the operators
 \ $a_N$ \ and
 \ $a^ + _N$ \ in the first power, but contains the contribution of a zero
  power which we shall designate by \ $e^ {a (0)} _i (x) $ \ .

  The information about configuration of the fields \ $e^a_ {Ni} (x) $ \ can
 be received by study the matrix elements of some invariant operators
 relative to the states (3.3). For example,
  consider the quantity \ $V=\int\, d^3x\,\sqrt {-g} $ \ which is invariant
relative to general  coordinate transformations in 3-D space. As
it is known, the expression in the right hand side of Eq. (2.8)
gives the change of tetrads under the
infinitesimal transformation. On the other hand, the right hand side
of Eq.(2.8) is equal to PB of tetrads and hamiltonian.
Since the hamiltonian annulates the
physical states, there is the equality
$$
 \langle\,0\,\vert\, (\,\int\, d^3x\,\sqrt
{ -g} \,) \,\vert\, N\,\rangle= \langle\,0\,\vert\,\, e^ {-i\varepsilon
 H} \, (\,\int\, d^3x\,\sqrt {-g} \,) \, e^ {i\varepsilon H}
  \,\vert\, N\,\rangle
$$
From here we obtain the equation
$$
\langle\,0\vert\,:\{\,\int\, d^3x\,\sqrt
{ -g} \, g^ {ij} \, (\, e^a_j\nabla_i\xi_a + \nabla_i\xi_a\cdot
e^a_j\,) \,\} \,:\,\vert\, N\,\rangle=0 ,            \eqno(3.10)
$$
which is true  for any field \ $\xi^a (x) $ \ .

To advance further, we shall assume, that the Heisenberg tetrads and
connection fields can be expressed as formal series in
operators \ $a_N$ \ and \ $a^ + _N$ \ . For the tetrad fields
the beginning of this expansion is given by Eq. (3.9).
Then we have the expansion for the connection operator
accordin to Eq. (3.1).
Let designate by \ $\omega^ {ab (0)} _i (x) $ \ the
 zero power term in operators
\ $a_N$ \ and \ $a^+_N$ \ .
This term can be expressed through the
 field \ $e^ {a (0)} _i (x) $ \ with the help of Eqs. (2.8), (2.9).

Now from equality (3.10) the following conditions for fields \ $e^a_ {Ni} $
are obtained
$$
\nabla^ {(0)} _i\, (\sqrt {-g^ {(0)}} \, g^ {ij (0)} e^a_ {Nj} \,) =0
\eqno (3.11)
$$
The top index \ $ (0) $ \ means that the all operators and fields under
this index depend only on zero approximation of the tetrad and
connection fields \ $e^ {a (0)} _i$ \ and \ $\omega^ {ab (0)} _i$.

   Designate by \ $e^ {a (s)} _i (x) $ \ and \ $\omega^ {ab (s)} _i (x),
	 \ s=0,1,\ldots$ \ the contributions to tetrad and connection fields of
 the power \ $s$ relative to the creation and annihilation operators, so that
 $$
 e^a_i (x) =\sum^ {\infty} _ {s=0} \, e^ {a (s)} _i (x) \,,
 $$
 $$
 e^ {a (s)} _i (x) =\sum_ {N_1\ldots N_s} \, a_ {N_1} \ldots
 a_{N_s} \, e^a_ {N_1\ldots N_s\, i} (x) +
 $$
 $$
 \sum_ {M_1} \,\sum_ {N_1\ldots N_ {s-1}} \, a^ + _ {M_1} \, a_ {N_1} \ldots
 a_{N_ {s-1}} \, e^a_ {M_1; N_1\ldots N_ {s-1} \, i} (x) +
 \ldots \eqno (3.12)
 $$
 Here the creation and annihilation operators are normally ordered. The
  similar formulae take place for the connection field.
  All information about evolution in time of the
 system is contained in the fields \ $ e^a_ {N_1\ldots N_s\, i} (x) \,,
  \ \omega^ {ab} _ {N_1\ldots N_s\, i} (x) $ \ and
   so on , \ $s=0,1,\ldots$ \ .
  We shall designate by \ $\Phi_ {\cal N} (x) $ \ the totality of these
  fields.

Denote the group consisting of  elements \ $ \{S^a_b (x)\}$ \  by
\ $G$ \ . To each element of group \ $G$ \ corresponds
the transformation of the fields of tetrads and connections:
$$
e^{\prime a}_i (x) =S^a_b (x) e^b_i (x),
$$
$$
\omega^{\prime ab}_i (x) =S^a_c (x) S^b_d (x)\omega^{cd}_i (x) +S^a_c (x)\eta^{cd}
\partial_iS^b_d (x)
$$
The operator
$$
\chi (\omega_0)=\frac{1}{2}\,\int\, d^3x\,\omega^{ab}_0\chi_{ab},
\qquad\omega^{ab}_0\longrightarrow 0
$$
is the right invariant  vector field on the group \ $G$ \ transfering
the point \ $S^a_b (x) $ \ in the infinitely close point \ $S^a_b (x) -
\omega^a_0{ }_c (x) \, S^c_b (x) $ \ .
The vector field on the  group\  ${\cal G}$ \ , corresponding
 to the operator
$$
\phi (e_0)=\frac{1}{2\,\kappa^2}\,\int\, d^3x\, e^c_0\phi_c,\qquad
e^c_0\longrightarrow 0,
$$
generates the following shift of tetrad fields (see (2.3) and (2.8)):
$$
e^a_i\longrightarrow e^a_i+\nabla_ie^a_0
$$

Now it is easy to express the field \ $\pi^i_a (x) $ \ , conjugated with the tetrad field
\ $e^a_i (x) $  \ (see (3.2))  in  the  first  approximation relative to the
operators \  $a^+_N$ \ and \ $a_N$ \ . For thees it is necessary to complement
the set of fields \ $e^a_{Ni} (x) $ \  in (3.9) up to complete set  of orthonormal fields
 \ $\{e^a_{Ni} (x)\}$ .
In consequence the following formulae take place:
$$
\int\, d^3x\,\sqrt {-g^{ (0) }}\, {\bar e}^a_{Mi}g^{ij (0) }e_{Naj}=
\kappa^2q^2_M\,\eta_{MN}, \eqno (3.13)
$$
where \ $\eta_{MN}=0$ \  if  \  $M\neq N\,,\ \eta_{NN}=1$ \  or
\ $-1$ \ for the
space- or time-like field \ $e^a_{Ni}$ \ ,
accordingly  . Here \ $q^2_M$ \ is the normalising multiplier having dimension  of
lengths.
Except the condition of orthonormality (3.13) there is  also the condition of
completeness:
$$
\sum_N\,\kappa^{-2}q^{-2}_N\,\eta_{NN}\, e^a_{Ni} (x) {\bar e}^b_{Nj} (x)=
$$
$$
\frac{1}{\sqrt {-g^{ (0) }}}\, g^{ (0) }_{ij}\eta^{ab}\delta (x-y) -
\nabla^{ (0) }_i (x)\nabla^{ (0) }_j (y) D^{ab (0)} (x, y),
$$
$$
-\nabla^{ (0) }_i\sqrt {-g^{ (0) }}\, g^{ij (0)}\nabla^{ (0) }_jD^{ab (0)} (x, y)=
\eta^{ab}\delta (x-y)\eqno (3.14)
$$
The set of operators \ $a^+_N$ \
and \ $a_N$ \ also must be complemented so to have instead CR (3.5) the CR:
$$
[a_M,\, a^+_N\,]=\eta_{MN}\eqno (3.15)
$$
Taking into account formulas (3.9), (3.14) and (3.15), we obtain the
following representation for the operator \ $ \pi^i_a$ \  :
$$
\pi^i_a (x)=
i\, (\,\sqrt {-g^{ (0) }}\, g^{ij (0)}\,) (x)\,\{\,
\frac{1}{{\sqrt 2}\,\kappa^2}\,\sum_N\, q^{-2}_N\,
(\, a_Ne_{Naj} (x)-a^+_N{\bar e}_{Naj} (x)\,)+
$$
$$
\nabla^{(0)}_j\,\int \, d^3z\, D_a{ }^ {c^{(0)}} (x, z)\frac{\delta}{\delta\xi^c (z)}\}
\eqno (3.16)
$$
The second term in (3.16) is vector field on  the group \ ${\cal G}$ \ , so
that
$$
[\,\int\, d^3z\, e^c_0 (z)\frac{\delta}{\delta\xi^c (z)}\, ,\ e^a_i
(x)\,]= \nabla_ie^a_0 (x)\eqno (3.17)
$$

From formulas
 (3.1), (3.9) and (3.14) - (3.17) it is seen that\\
1) in the first approximation relative to
 operators \ $a^+_N$ \ and \ $a_N$ \ the fields of tetrads  and connections
satisfy to quantum PB (2.12) and (2.13); \\
2) the fields of connections
also, just like the fields of tetrads, in the first order  relative to the
operators \  $a^+_N$ \ and \ $a_N$ \ contain all these operators.

 It is clear from the end of this Section (beginning with (3.13)) that
the fields of tetrad and connection have more degrees of freedom, than it is necessary
from kinematic considerations. Really, except the degrees of freedom,
corresponding to the gauge group \ ${\cal G}$ \  and contained in  the fields
\ $\Phi_{\cal N} (x) $ \ , there is the overfuled system of the degrees of freedom,
contained in the set of all creation and annihilation operators .
However, this  does not prevent  further advancement by  the  following
 reasons:
under regularization almost all the creation and annihilation operators are eliminated
and in the theory  only minimal their number  remains. Thus  under regularization
the gauge group  remains  intact.

Now we have all necessary means for regularization of the
theory.
\centerline{ }
\centerline{ }
\centerline{ }
\centerline{ }

\centerline{{\bf {4. Regularization. }}}
\centerline{ }

The description in previous Section carried the formal character, since
the divergences were not taken into account . In this Section the regularization of the theory is carried out.
 Thereby  the mathematical
sense is imparted to all used operators and equations.

The importance of CR (3.5) and (3.6) is that {\underbar { every}}   set of pairs of the
operators \ $a_N, \ a^ + _N$ \ can be considered as a set of the second
class constraints in the Dirac sense [3].  It enables the following realization of
regularization.

Select a finite set of pairs of annihilation and creation  operators
\ $\{ a_N,\, a^ + _N\}^{\prime} $ \ and
numerate them in such a manner that \ $ | N\, | < N_0$. The operators from the set
\ $\{a_N,\, a^+_N\,\}^{\prime}$ \  satisfy CR (3.5).
 Since the
physical information is contained in wave functions \ $e^a_ {Ni}(x) $ \ ,
 this choice is actually defined by the choice of a set of
linearly independent wave functions \ $\{ e^a_{Ni}(x) \}^{\prime} $ \, ,
corresponding to the set of operators \ $\{ a_N,\, a^+_N\}^{\prime} $ \ .
The choice of functions in the set \ $\{ e^a_{Ni} (x) \} ^ {\prime } $ \  is
determined by physical conditions of a problem. For example, if \ $x$-space is
torus then as the wave functions of this set the periodical  waves with wave numbers
 limitted  by modulus can be taken.
Such choice of  the set $\{\, e^a_{Ni} (x)\,\}^{\prime}$ \ corresponds to  the problem
of  gravitational waves.

The regularization of the theory consists in the
following: the all infinite number of pairs of the annihilation and creation
operators  at \ $\vert\, N\,\vert > N_0$ \ , i.e. except chosen,
we believe equal to zero:
$$
a_N=0, \qquad \ a^ + _N=0, \ \qquad \vert\,
N\,\vert > N_0 \eqno (4.1)
$$
Thereby  the infinite set of the second class constraints is imposet. Now we must
replace CR (2.12), (2.13) et al. to  corresponding  Dirac CR and
investigate  the  regularized equations of motion.

We shall prove the important theorem, which
made sensible the all schema of dynamic quantization: \\
\centerline { }

{ \underbar { Theorem }}.
{ \it { The imposition of the second class constraints (4.1) does not
change the form of Heisenberg equations, saving their classical form. }} \\
\centerline { }

Proof.

Let \ $\vert\, { \cal M } ^ { \prime } \,\rangle,\, \vert\, { \cal N } ^ {
\prime } \,\rangle,\ldots$ \ designate the basic vectors (3.3) constructed
with the help of the regularized set of operators
 \ $\{ a_N,\, a^+_N\}^{ \prime } $, \ and \ $F^{ \prime } $ \ designate the
 Fock space with these basic vectors. The imposition of
constraints (4.1) means that the space of physical states
 \ $F$ \ is reduced to the regularized subspace
\ $F^{\prime} \subset F$ \ . By definition for any operator \ $A$ \ in
the regularized theory only the matrix elements
\ $\langle\, { \cal M } ^ { \prime } \,\vert\, A\,\vert\, { \cal N } ^ {
\prime } \,\rangle$ \ are considered and the
 operators \ $a_N$ \ and \ $a^+_N$ \ with \ $\vert\, N\,\vert > N_0$
 \ contained in the operator \ $A$ are put  equal to zero after the normal
 ordering. Therefore, in regularized theory the matrix elements of quantum
DB for operators \ $A$ \ and \ $B$ \ corresponding to the constraints (4.1)
are represented in the form $$
\langle\, { \cal M } ^ { \prime } \,\vert\, [ A,\, B\, ] ^ * \vert\, { \cal
N } ^ { \prime } \,\rangle= \sum_ {{ \cal L } ^ { \prime }} \, (
\,\langle\, { \cal M } ^ { \prime } \,\vert\, A\, \vert\, { \cal L } ^ {
\prime } \,\rangle\,\langle\, { \cal L } ^ { \prime } \,\vert\, B\,\vert \,
{ \cal N } ^ { \prime } \,\rangle-
$$
$$
\langle\, { \cal M } ^ { \prime }
\,\vert\, B\, \vert\, { \cal L } ^ { \prime } \,\rangle\,\langle\, { \cal L
} ^ { \prime } \,\vert\, A\, \vert\, { \cal N } ^ { \prime } \,\rangle\, )
\eqno (4.2)
$$
By  definition of quantum DB the
 operators \ $a_N$ \ and \  $a^+_N$ \ with
\ $\vert\, N\,\vert > N_0$ contained in the operators \ $A$ \ and \  $B$ \
from (4.2) are normally  ordered and then put  equal to zero.  In contrast
to DB (4.2) in PB \ $ [ A,\, B\, ] $ \ at the calculation of matrix
elements \ $\langle\, {\cal M}^{\prime}\,\vert\, [A,\, B\,]\,\vert\, {\cal
N}^{\prime}\,\rangle$ \  by the formula, similar  to (4.2), the summation
 goes over  all  intermediate states (3.3).  Suppose that
the operator \ $B$ \ is diagonal in basis (3.3) and does not depend on the
operators \ $a_N$ \ and \ $a^ + _N$ \ with \ $\vert\, N\,\vert > N_0$.
Then it is seen from (4.2) , that
$$
\langle\, { \cal M } ^ { \prime }
\,\vert\, [ A,\, B\, ] ^ * \, \vert\, { \cal N } ^ { \prime } \,\rangle \ =
  \langle\, { \cal M } ^ { \prime } \,\vert\, [ A,\, B\, ] \,\vert\, { \cal
  N } ^ { \prime } \,\rangle \ , \eqno (4.3)
$$
if in a right hand side of Eq.(4.3) the
 operators \ $a_N \ , \ a^+_N$ \ at \ $\vert\,
N\,\vert > N_0$ \  in operator \ $A$ \ are normally ordered and then
are put equal to zero.
The CR (3.6 ) imply that the
Hamiltonian of the theory does not depend on the operators \ $a_N, \ a^ +
_N$ \ .  Therefore in (4.3) it is possible to substitute
 Hamiltonian \ $H$ \ for operator \ $B$ \ .  It means that the theorem is
true.

There is also classical variant  of  the Theorem.\\

{\it {The imposition of the second class constraints (4.1) does not change classical form of
Hamilton equations  for the remained degrees of freedom. }}
\centerline{ }

For provement  we shall write out the formula
 for  Dirac bracket in classic theory.

Let \  $\{\chi_{\alpha}\}$ \ and \ $\{\kappa_n\}$ \ designate the finite or
 infinite  sets of the first  and the second
  class constraints respectively.  By
definition this means, that
$$
[\,\chi_{\alpha},\,\chi_{\beta}\,]\approx 0,\eqno (4.4)
$$
$$
[\,\chi_{\alpha},\,\kappa_n\,]\approx 0\eqno (4.5)
$$
$$
[\,\kappa_m,\,\kappa_n\,] =c^{-1}_{mn}\eqno (4.6)
$$
Following to Dirac, we denote by the symbol \ $\approx$ \ the equalities  modulo
terms,which are proportional to constraints \ $\kappa_n$ \ or \ $\chi_{\alpha}$ \ .
We shall pay attention that the matrix \ $c^{-1}_{mn}$ \  in (4.6) is
nondegenerate matrix which in general case  is  dependent on dynamical
variables. The Hamiltonian of the system \ $H$ \ is  the first  class quantity
$$
[\, H,\,\chi_{\alpha}\,]\approx 0,\eqno (4.7)
$$
$$
[\, H,\,\kappa_n\,]\approx 0\eqno (4.8)
$$
In  classic  theory the Dirac bracket of   any two quantities   is  defined
by the formula
$$
[\,\xi,\,\eta\,] ^*= [\,\xi,\,\eta\,] -\sum_{m, n}\, [\,\xi,\,\kappa_m\,]\, c_{mn}\,
[\,\kappa_n,\,\eta\,]\eqno (4.9)
$$
Evidently, for any quantities \ $\xi$ \ and \ $\kappa_n$ \ we have
$$
[\,\xi,\,\kappa_n] ^*=0
$$
From here it follows that the second class constraints \ $\kappa_n$ \ can be
put equal to zero before the calculation of Dirac brackets.
Because of  (4.8) and (4.9)  the  equation
$$
[\,\xi,\, H\,]\approx [\,\xi,\, H] ^*\eqno (4.10)
$$
is valid.
The weak Eq.(4.10)
means, that the equations of motion obtained with the help of the Poisson brackets and
Dirac brackets, essentially coincide.

In the context of our method according to (3.6) the weak equalities (4.5) and (4.8)
are transformed in strong equalities. That is why from (4.9)  it immediately  follows, that
\ $ [\,\xi,\, H\,]=[\,\xi,\, H] ^* $ \ for any quantity \ $\xi$ \ . It
means that the theorem is true.\\
\centerline{ }

{ \underbar { Consequence. }} { \it { The regularized theory is
general covariant. }} \\
\centerline { }

This Consequence directly follows from the proved theorem.
Indeed, the equations of motion in regularized theory, coincide in the form with
 classical one which are general covariant.

We call our quantization method as dynamic namely for the reason, that  in this method the regularization is ideally agreed
with dynamics of the system.
Once again  we pay attention to the fact, that the regularization does not touch the
gauge group  of the theory i.e. the  group  \  ${\cal G}$ \  in regularized theory remains  such,
 which it is in classical theory.

Now we shall state more formal approach to the dynamic quantization method.
This approach, being perhaps less natural, is more logically harmonic
 and also permit to simplify some calculations.

The basis of such approach is

{ \underbar { Assumption. }} { \it { The theory is
assumed regularized so that the following
axioms are true: }}
\\ \centerline{ }

{ \underbar { Axiom R1. } { \it { All states of the theory, having the
physical sense, are obtained from the ground state }} \ $\vert
\,0\,\rangle$ \ { \it { with the help of the creation operators }}
 \ $a^ + _N$ \ with \ $\vert\, N\,\vert < N_0$ \ :
 $$
 \vert \, n_1,\,
 N_1;\ldots;\, n_s,\, N_s\,\rangle= ( n_1!\cdot\ldots\cdot n_s!\, ) ^ {
 -\frac { 1 } { 2 }} \cdot ( a^ + _ { N_1 } ) ^ { n_1 } \cdot\ldots\cdot (
 a^ + _ { N_s } ) ^ { n_s } \,\vert \,0\,\rangle \ ,
 $$
 $$
 a_N\,\vert\,0\,\rangle=0                              \eqno (4.11)
 $$
 { \it
 { The states (4.11) form orthonormal basis of the physical states
  space \ $F^{\prime}$ \ of the theory. }} \\
\centerline{ }

{\underbar { Axiom R2. }}
{\it {The states  (4.11) satisfy  the  conditions }}
$$
\chi_{ab} (x)\,\vert \ \rangle=0,\qquad\phi_a (x)\,\vert \ \rangle=0
\eqno (4.12)
$$
\\
\centerline{ }

{ \underbar { Axiom R3. }}
{ \it { The dynamic variables}} \ $e^a_i(x)$  \ { \it { transfer the state (4.11) into some
 superposition of states  of  the  theory.  This  superposition
contains the { \bf all } states for which one of occupation number differs
 per unit from corresponding occupation number of the state (4.11) and the rest
 occupation numbers coincide with
 corresponding occupation numbers of the state  (4.11). }} \\
 \centerline { }

{ \underbar { Axiom R4. }}
{ \it { The equations of motion and constraints for
physical fields \ $e^a_i(x),
\omega^{ab}_i(x)$ \ coincide in form (up to arrangement of the
operators) with the corresponding classical equations  of motions and
constraints .}} \\
\centerline{ }

The Axioms R1 - R3 are analogous to Axioms 1- 3 in
the nonregularized theory. The Axiom R4 replaces the Theorem.
 It postulates a correct form of equations of motion and
constraints in agreement with classical mechanics.

Finally let us discuss briefly a serious problem of operators ordering in
regularized Heisenberg equations in general form.
The fields  \  $e^a_i (x) $\ and \ $\omega^{ab}_i (x) $ \ will be designated
 by one symbol \ $\Phi (x) $ \, and the fields \ $e^a_0 (x) $ \  and
\  $\omega^{ab}_0 (x) $ \ by the symbol \ $\lambda (x) $ \ .
The problem of operators ordering arises when we find out
the selfconsistency of the theory [3].  Let us write out the Heisenberg
equation in general form:
$$
{\dot {\Phi}}(x)=F(\lambda,\,\Phi\,)(x) \eqno(4.13)
$$
Here the field \  $\lambda(x)$ \ determines the Hamiltonian according to
(2.3).  In (4.13) \ $F$ \  is a local function of the fields
 \ $\lambda(x)$ \  and \ $\Phi(x)$ \  and linearly depends on the field
  \ $\lambda$ \ . By definition the
   field \ $\lambda$ \ is placed at the left of the fields \ $\Phi$ \ in
function \ $F(\lambda,\,\Phi\,)$ \ . According to (4.13) we have for any
 fields \ $\lambda_1$ \  and \ $\lambda_2$ \ :
 $$
\delta_i\Phi=\delta t_i\,F(\lambda_i,\,\Phi\,)\, , \qquad i=1,2
$$
Consider the quantity \ $(\delta_1 \delta_2-\delta_2\delta_1\,)\,\Phi$ \
which is denoted by \ $\delta_{[1\,2]}\Phi$ \ .  The necessary
  condition of the selfconsistency of the theory is the possibility of a
 such operators ordering in Eq.(4.13) that the weak equality
 $$
 \delta_{[1\,2]}\Phi\approx \delta t_1\,\delta
 t_2\,F\,(\lambda_{[1\,2]},\,\Phi\,)  \eqno(4.14)
 $$
 takes place. Here the field \ $\lambda_{[1\,2]}$ \ is a bilinear
 antisymmetric form of the fields \ $\lambda_1$ \  and \ $\lambda_2$ \  and
  it generally depends on the field \ $\Phi$. The first class
   constraints \ $\chi_{ab}$ \ and \ $\phi_c$ \ , which are vector fields,
on the group \ ${\cal G}$ \ , do not contain the creation and annihilation
 operators \ $ a^+_N, \ a_N $ \  .  The last  property  obviously
is used in calculations in the next Section.

\centerline{ }
\centerline{\bf {5. Perturbation  theory.}}
\centerline{}

In this Section we shall show, in what way
the field  coefficients at operators \ $a_N$ \ and \ $a^+_N$ \  in
 expansions of the tetrad and connection fields
  (see (3.12) can be find step by step.
 The calculations begin without quantum corrections (loops).
 Then  the result  is  corrected  with allowance for
quantum fluctuations. This   proves to be
 formally  equivalent to the expansion in the  number \ $N_0$ \ . As shown below,
the formal expansion in the  number \ $N_0$ \  is equivalent
to expansion in dimensionless parameter \ $ (\,\Lambda\,\kappa\,) ^2$ \ ,
where \ $\Lambda$ \  is the cut off momentum of  the theory.
If the cut off momentum much
less of Planckian  momentum, then
\ $ (\,\Lambda\,\kappa\,) ^2\ll 1$ \ .

The description in this Section is  very sketchy. The detailed
 study of PT and investigation of  concrete problems
must be carried out with the help of the dynamic method
in the  special woks.

Calculations are begin with zero
approximation \ $e^{a (0) }_i (x) $ \ and  \ $\omega^{ab (0) }_i (x) $ \ .
The  tetrad and connection fields  in zero approximation satisfy equations and
 constraints (2.8), (2.9) and (2.11) and they do not depend on
creation and annihilation operators. However, in the zero
 approximation tetrad and connection
 fields
are operators on the group of gauge
 transformations \ ${\cal G}$ \ according to
 (3.1), (3.16), (3.17).
Thus,  the fields  \ $e^{a (0) }_i (x) $ \ and
\  $\omega^{ab (0) }_i (x) $ \ satisfy equations of motion (2.8)
 and (2.10), and the constraints \ $\chi^{ (0) }_{ab}$ \
 and \  $\phi^{ (0) }_c$ \ composed from these fields,
by definition annulate the state \ $\vert\, 0\,\rangle$ \ .

In the first approximation the all quantum states
(4.11) from regularized space are involved in consideration. The tetrad and connection fields
are expanded
 in the first
approximation as follows (see (3.9) and (3.12)):
$$
e^a_i (x)=\frac{1}{\sqrt 2}\,\sum_{|N| < N_0}\, (\, a_Ne^a_{Ni} (x)+
a^+_N{\bar e}^a_{Ni} (x)\,)+
e^{a (0) }_i (x)\equiv e^{a (0) }_i (x) +e^{a (1) }_i (x),
$$
$$
\omega^{ab}_i (x)=\frac{1}{\sqrt
2}\,\sum_{|N| < N_0}\, (\, a_N\,\omega^{ab}_{Ni} (x) +a^+_N\,
{\bar {\omega}}^{ab}_{Ni} (x)\,)+\omega^{ab (0) }_i (x)\equiv
$$
$$
\equiv\omega^{ab (0) }_i (x)+\omega^{ab (1) }_i (x)
\eqno (5.1)
$$
We  remark, that
at our approach the field \ $ e^a_0$ \ and \   $  \omega^{ab}_0$ \ playing  the role  of
the Lagrange multipliers  remain
numerical:
$$
e^{a (s) }_0=0,\  \qquad
\omega^{ab (s) }_0=0, \qquad s=1, 2,\ldots \eqno (5.2)
$$
Substituting the fields from (5.1) in equations (2.8), (2.9) and (2.11) and
taking into account the equation (5.2) and that fact, that the
fields \ $e^{a (0) }_i\,,\ \omega^{ab (0) }_i$ \ satisfy all classical
equations, we obtain:
$$
\varepsilon_{abcd}\varepsilon_{ijk}\,\{\, R^{ab
(0) }_{ij}e^{c (1) }_k+ 2\, e^{c (0) }_k\nabla^{ (0) }_i\omega^{ab (1)
}_j\,\}=0,
$$
$$
\varepsilon_{ijk}\, (\,\nabla^{ (0) }_ie^{a (1) }_j-e^{ (0) }_{bi}
\omega^{ab (1) }_j\,) =0,\eqno (5.3)
$$
$$
\varepsilon_{abcd}\varepsilon_{ijk}\,\{\, R^{ab (0) }_{0i}e^{c (1) }_j+
e^{c (0) }_j\nabla_0\,\omega^{ab (1) }_i+ e^c_0\nabla^{ (0) }_i\omega^{ab
(1) }_j\,\}=0,
$$
$$
\nabla_0\, e^{a (1) }_i-e_{b0}\omega^{ab (1) }_i=0\eqno (5.4)
$$
Equations (5.3) are the constraints, and equations (5.4) are equations
of motion for the tetrad and connection fields. Taking the matrix elements
of  Eqs. (5.3) and (5.4) of the form
 \ $\langle\, 0\,\vert\ldots\vert\, N\,\rangle$ \ , we find
$$
\varepsilon_{abcd}\varepsilon_{ijk}\,\{\, R^{ab (0) }_{ij}e^c_{Nk}+ 2\,
e^{c (0) }_k\nabla^{ (0) }_i\omega^{ab}_{Nj}\,\}=0, \eqno (5.5 a)
$$
$$
\varepsilon_{ijk}\, (\,\nabla^{ (0) }_ie^a_{Nj}-e^{ (0) }_{bi}
\omega^{ab}_{Nj}\,) =0,\eqno (5.5 b)
$$
$$
\varepsilon_{abcd}\varepsilon_{ijk}\,\{\, R^{ab (0) }_{0i}e^c_{Nj}+
e^{c (0) }_j\nabla_0\,\omega^{ab}_{Ni}+ e^c_0\nabla^{ (0)
}_i\omega^{ab}_{Nj}\,\}=0,\eqno (5.6 a)
$$
$$
\nabla_0\, e^a_{Ni}-e_{b0}\,\omega^{ab}_{Ni}=0\eqno (5.6 b)
$$
Thus, the equations of  constraints  (5.5) and motions (5.6) break up
on separate  equations with given number \ $N$ \ .
The same property takes place for
fields \ $e^a_{N_1\ldots N_s\, i}$ \ and etc.,
if quantum fluctuations do not  taken into account. For example,  we
have the following analog of equations (5.5 b) and (5.6 b) for the fields
 \ $e^a_{N_1N_2\, i}$ \ and  \ \  $\omega^{ab}_{N_1N_2\, i}$ \ :
 $$
\varepsilon_{ijk}\,\{\,\nabla^{ (0) }_ie^a_{N_1N_2\, j}+e^{ (0) }_{bj}
\omega^{ab}_{N_1N_2\, i}+\omega^{ab}_{N_1\, i}e_{N_2\, bj}+
\omega^{ab}_{N_2\, i}e_{N_1\, bj}\,\}=0\,,\eqno (5.7)
$$
$$
\nabla_0e^a_{N_1N_2\, i}-e_{b0}\omega^{ab}_{N_1N_2\, i}=0\eqno (5.8)
$$
It is seen , that in opposite  to
 Eqs.(5.5) and (5.6) , which are uniform relative to the fields
 \ $e^a_{N\, i}$ \ and \ $\omega^{ab}_{Ni}$ \ ,
Eqs.(5.7) are not uniform relative to the fields
\ $e^a_{N_1N_2\, i}$ \ and \ $\omega^{ab}_{N_1N_2\, i}$ \ .

To solve Eqs. (5.5) - (5.8) it may be applied the following  scheme.
At first it is necessary  to solve the uniform system of  Eqs. (5.5) and (5.6)
relative to the fields  \ $e^a_{Ni}$ \ and \ $\omega^{ab}_{Ni}$\  . Then we solve
the linear nonuniform system of equations including Eqs. (5.7) and
(5.8) relative to the
 fields  \ $e^a_{N_1N_2\, i}$ \ and \ $\omega^{ab}_{N_1N_2\, i}$ \ .
This system of equations is dependent on  the fields
 \ $e^a_{Ni}$ \ and \ $\omega^{ab}_{Ni}$ \  found before. Further this
process spreads  on the higher fields  \ $e^a_{N_1\ldots N_s\, i}$ \ and
etc. Without  the quantum fluctuations  the arising  equations relative
  to the fields \ $e^a_{N_1 \ldots N_s\, i}$ \ and etc. are
linear nonunoform equations dependent on the fields  found on previous  steps.

The number  of constraints  and equations (5.5) and (5.6) is equal to  fourty,
and  number of the unknown
functions \ $e^a_{Ni}$ \ and \  $\omega^{ab}_{Ni}$ \  (at fixed index
\ $N$ \ ) is equal  to thirty. Nevertheless
the system of Eqs.(5.5) and (5.6) has
nonzero solutions. Really, the equations of motion (5.6) have  the solutions at
any  meanings of fields  \ $e^a_0$ \ and \ $\omega^{ab}_0$ \ . This is obvious for
initial  Eqs. (2.8), (2.9) and (2.11).
Therefore  the obtained from Eqs. (2.8), (2.9) and (2.11)
the equations
(5.5) and (5.6) relative to the
fields \ $e^{a}_{Ni}$ \ and \ $\omega^{ab}_{Ni}$ \ have  solutions.

Solving Eqs.(5.5) and (5.6) one must use  the initial
 conditions (3.11)
for the fields \ $e^a_{Ni}$ \ at \ $t=t_0$ \ .  The field \ $\omega^{ab}_{Ni}$
\ is expressed unambiguously through the
 fields  \ $e^a_{Ni}$ \ and \ ${\dot e}^a_{Ni}$ \ with the help  of  Eqs.
 (5.5 b) and (5.6 b). Then equations (5.5 a) and (5.6 a) result in linear
differential equations  of the second order relative to
 the fields \ $e^a_{Ni}$ \ .

From normalization condition (3.13) and equations (5.5) and (5.6) it is seen, that the
fields \ $e^a_{Ni}$ \ and \ $\omega^{ab}_{Ni}$ \ are proportional to the gravitational
constant \ $\kappa$ . Now  become clear the sense of multiplication
on constant \ $\kappa^2$ \  in  the right hand side of normalization  condition (3.13) .
The expansion (5.1) and proportionality of the fields \ $e^a_{Ni}$ \ and
\ $ \omega^{ab}_{Ni}$ \ to the  constant \ $ \kappa $ \  gives the possibility of fulfilment
of the Poisson brackets (2.13) in lowest approximation relative to
the creation and annihilation operators (in nonregularized theory).

Thus, the fields  \ $e^a_{Ni}$ \ and \ $ \omega^{ab}_{Ni}$ \  are found
according to
the following  rule. One must find the  fields  \ $e^a_{Ni}$ \ and
 \   $ \omega^{ab}_{Ni}$ \ ,
constraind to the conditions (3.11) at \ $t=t_0$ \  , (3.13) and equations (5.5)
and (5.6). The tetrad and connection  fields  composed  from
found fields according to (5.1) (where summation goes over all
 \ $N$ \ ), must satisfy the Poisson brackets
(2.13) in lowest наинизшем approximation. Then from the  all set of
the  fields \ $e^a_{Ni}$ \ and \  $\omega^{ab}_{Ni}$ \ we choose
the regularized
 subset \  $\{\, e^a_{Ni},\,\omega^{ab}_{Ni}\,\}^{\prime}$ \ with \ $|N| < N_0$ .

We  notice, that according to equations (5.7) we have
$$
e^a_{N_1N_2\, i}\sim\kappa^2\,,\qquad\omega^{ab}_{N_1N_2\, i}\sim
\kappa^2\eqno (5.9)
$$

Consider shortly the question about  quantum fluctuations or loops.

Denote  by \ $e^{a (s) }_{Ni},\ s=0, 1,\ldots$ ,\ the  $s$-loop contributions in the fields
\ $e^a_{Ni}$ \ .
Thus  the considerred  above  fields
\ $e^a_{Ni}$ \ correspond to the fields \ $e^{a (0) }_{Ni}$ \ in new
notations.

By the same way , as equations (5.5) - (5.8) were obtained,  we find the
following equations in one-loop approximation:
$$
\varepsilon_{ijk}\,\{\,\partial_ie^{a (1) }_{Nj}+\sum_{|M| < N_0}\, [\, 2\, (\,
\omega^{ab (0) }_{NMi}\, {\bar e}^{ (0) }_{Mbj}+{\bar {\omega}}^{ab (0) }_{Mi}\,
e^{ (0) }_{NMbj}\,)+
$$
$$
+(\,\omega^{ab (0) }_{M; Ni}\, e^{ (0) }_{Mbj}+
\omega^{ab (0) }_{Mi}\, e^{ (00) }_{M; Nbj}\,)
\,]\,\}=0\eqno (5.10)
$$
From dimensional considerations (see (5.9)) it is  easy to understand, that the sum in
last equation
is  of the  order \ $ (\,\Lambda\,\kappa\,) ^2$ \ , where \ $\Lambda$ \ is
the cut-off momentum of  the theory.
The expansion at successive accounting of quantum fluctuations goes
namely in dimensionless parameter
\ $ (\Lambda\kappa\,)^2$ \  . Now the question about conception "small
number" of the physical degrees of freedom is clearified. This
 is such a number, at which
$$
(\Lambda\kappa)^2\ll 1 \eqno (5.11)
$$
Condition (5.11) means, that the cut-off momentum is much  less, than
Planckian   momentum. In this case the account of quantum fluctuations can
be made  with the help  of finite PT so , as shown above.

\centerline{ }
\centerline{\bf {6. Conclusion }}
\centerline{ }

Thus,  we described the new scheme of
canonical quantization of gravity.  The constucted quantum
theory
has following basic properties.\\

A. If  \ $x$-space is compact than the number of the physical
 degrees of freedom is finite.

B. The Heisenberg equations for tetrad and connection
and other fields are of
classical form (up operators ordering).

C. The constucted theory is general covariant.\\
\centerline{ }

Unfortunately, we have to do an essential reservation.
The mathematical
 correctness of the theory will be completely established only,
when the problem of operators  ordering in equations of
motions will be solved.  In this paper the mathematical correctness
 is established for  decided  here problems of
ultraviolet divergences
and general covariance of the theory. On the first sight
 the problem of creation and annihilation operators ordering
in Section 5 is resolved automatically, since the
coefficients at the creation and annihilation operators
in equations of motion
and constraints are put equal to zero. Thus we obtain
the equations for
the fields \ $\Phi_{{\cal N}}$ \ .
However, in this case  the question about
correctness of these equations arises. This question needs
special examination.

Pay  our  attention, that the condition (5.11)  of existence of PT
is not necessary for the mathematical correctness of the theory.
In our
opinion,  the physically sensible quantum theory
of gravity should not  be restricted by condition (5.11).
Nevertheless, it is possible, that in some
concrete examples the condition (5.11) will effectively take place.

Though in this paper the theory
 was built  in the case  of  pure
gravity, the inclusion of matter in the theory on the first  sight
cannot  lead to
principle difficulties under dynamical quantization.
In this direction the most interest  for us presents  the studies
of supergravity, since  the property of supersymmetry of
the theory is easier  established on the equations of motion
 playing the main
role in dynamic quantization  method.\\

\centerline{ }
\centerline{\sl {References. }}
\centerline{ }

\begin{itemize}
\item [1.]
Vergeles S. N. ,  Zh. Eksp. Teor. Fiz., 102 (1992) 1739.
\end{itemize}
\begin{itemize}
\item [2.]
Vergeles S. N. , Yad. Fiz., 57 (1994) 2286.
\end{itemize}
\begin{itemize}
\item [3.]
Dirac P. A. M. Lectures on Quantum Mechanics.
N. Y.: Yeshiva Univ., 1964.
\end{itemize}
\begin{itemize}
\item [4.]
Ashtekar A. and Isham C. J. , Class. Quantum Grav. , 9
(1992) 1433;\\
Ashtekar A. , Mathematics and General Relativity. ,
 AMS, Providence 1987;\\
Rendall A. , Class. Quantum Grav., 10 (1993)  605;\\
Ashtecar A. and Lewandowski J. , J. Math. Phys. , 36
(1995)  2170;\\
Ashtekar A., Lewandowski J., Marolf D.,
Mourao J. and Thiemann T. ,
J. Math. Phys. , 36 (1995). 6456;\\
Rovelli C. and Smolin L. , Nucl. Phys. B. , 442 (1995) 593;\\
Rovelli C. , Nucl. Phys. B. , 405 (1993) 797;\\
Smolin L. , Phys. Rev. D., 49 (1994) 4028;\\
Carlip S., Class. Quantum Grav., 8 (1991)  5;\\
Carlip S., Phys. Rev. D. , 42 (1990)  2647.
\end{itemize}
\begin{itemize}
\item [5.]
Xiang X. ,  Gen. Rel. Grav., 25 (1993) 1019.;\\
Ashtekar, A., Balachandran, A. P., and Jo. S. ,
 Int. J. Mod. Phys. A. , 4 (1989) 1493.
\end{itemize}
}}}

\end{document}